\documentclass[twocolumn,showpacs,preprintnumbers,amsmath,amssymb]{revtex4}

\usepackage{graphicx}
\usepackage{dcolumn}
\usepackage{bm}

\begin{document}

\title{Kaonic hydrogen atom and $K^-p$ scattering length}
\author{Y.~Yan}
\email{yupeng@sut.ac.th}
\affiliation{School of Physics, Institute of Science, Suranaree University of Technology,
Nakhon Ratchasima 30000, Thailand \\
Thailand Center of Excellence in Physics, Ministry of Education, Bangkok, Thailand}
\date{\today}

\begin{abstract}
Kaonic hydrogen is studied with various realistic potentials in an accurate
numerical approach based on Sturmian functions. The $K^-p$ scattering
length extracted from the $1s$ energy shift of the kaonic hydrogen
by applying the Deser-Trueman formula is severely inconsistent with
the one derived by directly solving the scattering Sch\"odinger equation.
We pay special attention to the recent measurement of the energy shift and decay
width of the $1s$ kaonic hydrogen state by the DEAR Collaboration.
After taking into account the large discrepancy between the extracted
and directly-evaluated scattering lengths, we found theoretical
predictions of most chiral SU(3) based models for the kaonic hydrogen decay width
are consistent with the DEAR data.
We warn the SIDDHARTA collaboration
that it may not be reasonable to extract kaon-nucleon scattering lengths,
by using the Coulomb-interaction corrected Deser-Truemab formula,
from the planned measurement of kaonic hydrogen.
\end{abstract}


\maketitle

Kaonic hydrogen is mainly the Coulomb bound state of a $K^-$ and a
proton but is affected by the strong interaction at small distances.
Furthermore, the strong interaction couples the $K^-p$ state to the
$\bar K^0 n$, $\pi\Sigma$, $\pi\Lambda$, $\eta\Sigma$ and
$\eta\Lambda$ channels and results in the $\pi\Sigma$ and
$\pi\Lambda$ decaying modes. It is believed that the study of kaonic
hydrogen effectively probes the low-energy, especially zero energy
strong kaon-nucleon interaction. Inspired by the recent precise
determination of the energy and decay width by the DEAR
Collaboration \cite{DEAR}, kaonic hydrogen has been extensively
studied in the theoretical sector, mainly in effective field theory
\cite{Oller2001,Oset2002,Hyodo2003,Meissner2004,Ivanov2004,Ivanov2005,Borasoy2005,Oller2005,Borasoy2006}.
Theoretical predictions for the $K^-p$ scattering length have
been compared with the DEAR data of the energy and decay width of
kaonic hydrogen, by using the
Deser-Trueman formula \cite{Trueman,Deser}
\begin{equation}\label{eqn:1}
\Delta E_{1s}+i\frac{\Gamma_{1s}}{2}=2\alpha^3\,\mu^2\,a_{p}
\end{equation}
or the Coulomb-interaction corrected Deser-Trueman formula \cite{Meissner2004}
\begin{equation}\label{eqn:2}
\Delta E_{1s}+i\frac{\Gamma_{1s}}{2}=2\alpha^3\,\mu^2\,a_{p}
\left[1-2\alpha\,\mu\,a_{p}\,(\ln\alpha-1)\right]
\end{equation}
where $\mu$ is the reduced mass of the $K^-p$
system, $\Delta E_{1s}$ and $\Gamma_{1s}$ are the energy shift and
decay width of the 1s kaonic hydrogen due to the strong interaction,
and $a_{p}$ stands for the S-wave $K^-p$ scattering length.
A general result is that the energy shifts and decay widths extracted,
by using eq. (\ref{eqn:1}) or (\ref{eqn:2}), from  theoretical
$K^-p$ scattering lengths are much larger than the DEAR data.

In this
work we show that eqs. (\ref{eqn:1}) and (\ref{eqn:2}) may not well
hold for
the $\overline KN$ system and the discrepancy between the theoretical
results and the DEAR data is not that large and particularly the theoretical
decay widths are consistent with the DEAR data.

We derive the $K^-p$
scattering length and kaonic hydrogen energy shift and decay width
by solving the same dynamical equation,
\begin{equation}\label{eqn:3}
\left[-\frac{1}{r^2}\frac{d}{dr}\left(r^2\frac{d}{dr}\right)+\frac{l(l+1)}{r^2}
-{\bf Q}^2+{\bf f}\,{\bf V}\right]{\bf R}(r)
\end{equation}
with
\begin{equation}
{\bf Q}^2=\left(
\begin{array}{cc}
q_c^2 & 0 \\
0 & q^2_0
\end{array}
\right),
\;\;
{\bf f}=\left(
\begin{array}{cc}
f_c & 0 \\
0 & f_0
\end{array}
\right),
\end{equation}
\begin{equation}
{\bf V} ={\bf V}^{em}+ {\bf V}^h,
\end{equation}
\begin{equation}
{\bf V}^{em}
=\left(
\begin{array}{cc}
V^{em} & 0 \\
0 & 0
\end{array}
\right),
\end{equation}
\begin{equation}
{\bf V}^h=
\left(
\begin{array}{cc}
\frac{1}{2}(V_1^h+V_0^h) & \frac{1}{2}(V_1^h-V_0^h)  \\
\frac{1}{2}(V_1^h-V_0^h)  & \frac{1}{2}(V_1^h+V_0^h)
\end{array}
\right),
\end{equation}
\begin{equation}
{\bf R}(r)=\left(
\begin{array}{c}
R_{K^-p}(r) \\
R_{\bar K^0 n}(r)
\end{array}
\right),
\end{equation}
\begin{equation}
q^2_c=\frac{[E^2-(M_p-M_{K^-})^2][E^2-(M_p+M_{K^-})^2]}{4E^2},
\end{equation}
\begin{equation}
q^2_0=\frac{[E^2-(M_n-M_{\bar K^0})^2][E^2-(M_n+M_{\bar K^0})^2]}{4E^2},\\
\end{equation}
\begin{equation}
f_c=\frac{E^2-M_p^2-M_{K^-}^2}{E}, \\
\end{equation}
\begin{equation}
f_0=\frac{E^2-M_n^2-M_{\bar K^0}^2}{E}
\end{equation}
where $V^{em}$ is the electromagnetic potential, $V_0^h$ and $V_1^h$
are respectively the
isospin I=0 and 1 strong interactions of the $\overline KN$ system,
$R_{K^-p}(r)$ and $R_{\bar K^0 n}(r)$ are
respectively the $K^-p$ and $\bar K^0n$ parts of the radial wave function of
the $\overline KN$ system. Eq. (\ref{eqn:3}) embeds into the Schr\"odinger
equation the relativistic effect
and the mass difference between the $K^-p$ and $\bar K^0n$ components.
The relativistic modification of the Schr\"odinger equation to eq.(\ref{eqn:3})
has been discussed in the works \cite{Auvil,Pearce,Gibbs,Gashi1,Gashi2,Gashi3}.

We study kaonic hydrogen and kaon-nucleon scattering with the
phenomenological $\overline KN$ potential
taken from the
work \cite{AY1,AY2} and the various effective potentials which are
worked out
in the work \cite{Hyodo}. The interaction \cite{AY1,AY2}
is constructed by fitting the free
$\bar KN$ scattering data \cite{Martin}, the $KpX$ data of kaonic
hydrogen by the KEK Collaboration \cite{KEK} and the binding energy
and decay width of $\Lambda(1405)$, which is regarded as an isospin
$I=0$ bound state of $\bar KN$. Since the interaction gives one
molecular state $\Lambda(1405)$, it must be much stronger than the
strong pion-pion interaction.

In the work \cite{Hyodo} an effective local potential in coordinate space
is constructed such as the solution of the Schr\"odinger or
Lippmann-Schwinger equation with such a potential approximates
as closely as possible the scattering
amplitude derived from the full chiral coupled-channel calculation.
Several realistic chiral SU(3) based models \cite{Hyodo2003,Oset2002,Borasoy2005,Borasoy2006}
have been studied.

The accurate evaluation of energy shifts, decay widths and especially
wave functions of exotic atoms has been a
challenge to numerical methods \cite{Gashi2,Amirk}. An approach is
required, which is able to account accurately for both the strong
short-range interaction and the long-range Coulomb force. The
numerical approach based on Sturmian functions \cite{Stur} has been
found effective and accurate. In this work we use the numerical
method which has been carefully studied and discussed in
\cite{Stur,Yanatom,Yanpionium} to study kaonic
hydrogen.

Shown in Table \ref{table:1} are our theoretical results with
various realistic $\overline KN$ potentials. The energy shift
$\Delta E_{1s}$ and decay width $\Gamma_{1s}$
of the $1s$ kaonic hydrogen state are derived by solving eq. (\ref{eqn:3})
in the above mentioned Sturmian function approach \cite{Stur,Yanatom,Yanpionium}.
The negative energy shifts in Table \ref{table:1} mean
that the $1s$ energy level is effectively pushed up by the strong interaction
since there exists one deep bound state, the $\Lambda(1405)$.
The $K^-p$ scattering lengths $a_{K^-p}$ in Table \ref{table:1} are directly evaluated
from eq. (\ref{eqn:3}). Listed
in the last column of Table \ref{table:1} are the extracted $K^-p$ scattering
length $\tilde{a}_{K^-p}$ from the energy
shifts $\Delta E_{1s}$ and decay width $\Gamma_{1s}$ in Column 1 and 2 by applying the
Deser-Trueman formula of eq. (\ref{eqn:1}).

\begin{table}
\caption{Energy shift $\Delta E_{1s}$, decay width $\Gamma_{1s}$ of kaonic hydrogen and
$K^-p$ scattering length $a_{K^-p}$ are derived by directly
solving the dynamical equation eq. (\ref{eqn:3}). $\tilde{a}_{K^-p}$ are extracted from the
energy shifts and decay widths in Column 1 and 2 by applying the
Deser-Trueman formula of eq. (\ref{eqn:1}). Energy shifts and decay widths
are given in eV.} \label{table:1}
\begin{tabular}{cccccccc}
\hline
\hline
$V_{\overline K N}(r)$ & $\Delta E_{1s}$   & $\Gamma_{1s}/2$    & $a_{K^-p}$ [fm]   &   $\tilde{a}_{K^-p}$ [fm]  \\
\hline
AY \cite{AY2} & -384  & 144 &   -$1.012 + 0.499 i$  & -$0.934 + 0.348 i$ \\
HNJH \cite{Hyodo2003} & -336 &  324 &   -$0.778 + 1.084 i$ & -$0.815 + 0.785 i$ \\
ORB \cite{Oset2002}& -348 &  323 &   -$0.804 + 1.067 i$ &  -$0.845 + 0.784 i$ \\
BNW \cite{Borasoy2005} & -288 &  337 &   -$0.625 + 1.068 i$ &  -$0.700 + 0.818 i$ \\
BMN \cite{Borasoy2006} & -297 &  311 &   -$0.655 + 0.992 i$ &  -$0.721 + 0.755 i$ \\
\hline
\hline
\end{tabular}
\end{table}

It is clear that with the same interaction \cite {AY2,Hyodo} the
scattering length derived by directly solving the Schr\"odinger
equation in eq. (\ref{eqn:3}) is rather different from the one
extracted from the energy shift and decay width of the $1s$ kaonic hydrogen
by applying the Deser-Trueman
formula of eq. (\ref{eqn:1}). One finds from Table \ref{table:1} that for the
imaginary part the extracted scattering length is smaller by a factor of 20\% to 30\%
than the directly-derived scattering length with the same interaction.
Averaging over the results for all the five potentials we get the averaged factor
to be 0.26. The result implies that both the lowest Deser-Trueman
formula in eq. (\ref{eqn:1}) and the Coulomb interaction corrected one in
eq. (\ref{eqn:2}) may not well apply to the $\overline KN$ system. The Coulomb interaction
leads to a correction, less than 10\%, to the lowest Deser-Trueman
formula in eq. (\ref{eqn:1}). Hence one may argue that
for the $\overline KN$ system an extraction, by using the formulas
in eqs. (\ref{eqn:1}) or (\ref{eqn:2}), may not be accurate, if not say, unreliable.

It has been puzzling that the energy shifts and decay widths, extracted from
the scattering lengths derived in a number of chiral
SU(3) based models \cite{Oller2001,Borasoy2005,Oller2005,Borasoy2006}, are inconsistent
with the DEAR data of kaonic hydrogen. However, we find, after considering the large
discrepancy between the directly evaluated and the extracted scattering lengths,
that the inconsistence between the theoretical decay width and the DEAR data
is not that obvious.

Listed in the second and third columns of Table \ref{table:2} are respectively
the isospin 0 and 1 $\overline KN$ scattering lengths predicted by various
chiral SU(3) based models \cite{Oller2001,Borasoy2005,Oller2005,Borasoy2006,Ivanov2004,Ivanov2005} where the isospin
symmetry limit is applied. In the fourth column we give the corresponding $K^-p$
scattering lengths evaluated from the isospin based scattering lengths $a_0$ and $a_1$,
by using the formula
\begin{equation}\label{eqn:4}
a_{K^-p}=\frac{(a_0+a_1)/2+a_0a_1\,q_0}{1+(a_0+a_1)\,q_0/2}
\end{equation}
with
\begin{equation}
q_0=\sqrt{2\mu_0\Delta}
\end{equation}
where $\mu_0$ is the reduced mass of the $\bar K^0n$ system and $\Delta$ the mass
difference between the $\bar K^0n$ and $K^-p$ systems. The formula in eq. (\ref{eqn:4})
holds with a high accuracy. We found a $K^-p$ scattering
length evaluated with the formula in eq. (\ref{eqn:4}) has a difference about 1\%
from the one derived by directly solving the particle based dynamical equation in
eq. (\ref{eqn:3}) for the potentials listed in Table \ref{table:1}. The decay widths
$\Gamma_{1s}$ in the last column of Table \ref{table:2} are extracted from the scattering
lengths listed in the fourth column with the
formula
\begin{equation}\label{eqn:5}
\frac{\Gamma_{1s}}{2}=2\alpha^3\,\mu^2\,{\rm Im}\, a_{K^-p}
\left(1-R\right)
\end{equation}
with $R=0.35$. We choose $R=0.35$ based on the facts that the Coulomb interaction
correction to the lowest Deser-Trueman
formula is up to 10\% and on average over the five realistic $\overline KN$ potentials
the extracted $K^-p$ scattering length with the lowest Deser-Trueman
formula is smaller by a factor of 0.26 than the directly evaluated one.

Comparing to the DEAR data that
\begin{equation}
\frac{\Gamma_{1s}}{2}=
125 \pm 56\; ({\rm stat}) \pm 15\; ({\rm syst})\; {\rm eV}
\end{equation}
one finds from Table \ref{table:2} that most decay widths extracted from
the scattering lengths predicted by the chiral SU(3) based models are
in line with the DEAR data of the $1s$ kaonic hydrogen decay width. We emphasize
that the DEAR result of the $1s$ kaonic hydrogen decay width is
well consistent with Martin's scattering data which leads, by applying the
modified Deser-Trueman formula in eq. (\ref{eqn:5}), to a decay width
$\Gamma_{1s}=202\pm 20$ eV.

\begin{table}
\caption{Isospin based scattering lengths $a_0$ and $a_1$ are taken from the works
listed in the table. $a_{K^-p}$ are evaluated with the formula in eq. (\ref{eqn:4}) and
decay widths $\Gamma_{1s}$ are extracted from the scattering lengths $a_{K^-p}$ in Column 4
by using the modified Deser-Trueman formula in eq. (\ref{eqn:5}).
 Experimental uncertainties
of Martin's scattering data are not shown in the table.} \label{table:2}
\begin{tabular}{cccccccc}
\hline
\hline
Ref. & $a_0$ [fm]   & $a_1$ [fm]    & $a_{K^-p}$ [fm]   &  $\Gamma_{1s}/2$ [eV]  \\
\hline
\cite{Oller2001}
 & -$1.31+1.24i$  & $0.26 + 0.66i$ &   -$0.64 + 1.15i $  & 307 \\
\cite{Borasoy2005} & -$1.48 + 0.86i$ &  $0.57 + 0.83i$ &   -$0.78 + 0.95i$ & 254  \\
\cite{Oller2005}  & -$1.23+ 0.45i$ &  $0.98 + 0.35i$ &   -$0.49 + 0.48i$ &  128  \\
\cite{Borasoy2006} & -$1.45 +  0.85 i$ & $0.65 + 0.76i$ & -$0.74 + 0.93 i$ & 248  \\
\cite{Borasoy2006} & -$1.72 +  0.77 i$ & $0.09 + 0.76i$ & -$1.11 + 0.86 i$ & 229  \\
\cite{Borasoy2006} &  -$1.64 + 0.75i$ & -$0.06 + 0.57i$ &   -$0.63 + 0.42i$ &  206  \\
\cite{Ivanov2004} &  -$1.22 + 0.54i$ & $0.26 + 0.00i$ &   -$0.78 + 0.60i$ &  112  \\
\cite{Ivanov2005} &  -$1.50 + 0.66i$ & $0.50 + 0.04i$ &   -$1.06 + 0.77i$   & 160 \\
\cite{Martin} & -$1.70+0.68i$ &  $0.37 + 0.60i$ &   -$1.03 + 0.76i$   & 202 \\
\hline
\hline
\end{tabular}
\end{table}

We conclude that for all the realistic local potentials employed in the work,
the $K^-p$ scattering
length extracted from the energy shift and decay width of the $1s$
kaonic hydrogen
by applying the lowest and Coulomb interaction corrected Deser-Trueman formulas is severely inconsistent with
the one derived by directly solving the scattering Sch\"odinger equation.
After taking into account the large discrepancy between the extracted
and directly-evaluated scattering lengths, we found theoretical
predictions for the kaonic hydrogen decay width by most chiral SU(3)
based models are consistent with the DEAR data.
We warn the SIDDHARTA collaboration
that it may not be reasonable to extract kaon-nucleon scattering lengths,
by using the Coulomb-interaction corrected Deser-Truemab formula,
from the planned measurement of kaonic hydrogen.

\section*{Acknowledgments}
This work is supported in part by the Commission on Higher
Education, Thailand (CHE-RES-RG Theoretical Physics). The author would like to thank
Th. Gutsche (Institute for Theoretical Physics, T\"ubingen
University), M. Lutz (GSI) and T. Hyodo (Department of Physics, Tokyo Institute of Technology)
for fruitful discussions.


\begin{thebibliography}{}
\bibitem{DEAR}
G. Beer {\it et al.} [DEAR Collaboration], Phys. Rev. Lett. {\bf 94}, 212302 (2005);
M. Gargnelli {\it et al.} [DEAR Collaboration], Int. J. Mod. Phys. A {\bf 20}, 341-348 (2005).

\bibitem{Oller2001}
  J.~A.~Oller and U.-G.~Mei{\ss}ner,
  Phys.\ Lett.\ B {\bf 500} (2001) 263
  [arXiv:hep-ph/0011146].

\bibitem{Oset2002}
E. Oset, A. Ramos, C. Bennhold,
Phys. Lett. B {\bf 527} (2002) 99
[arXiv:nucl-th/0109006].

\bibitem{Hyodo2003}
T. Hyodo, S.I. Nam, D. Jido, A. Hosaka,
Phys. Rev. C {\bf 68} (2003) 018201
[arXiv:nucl-th/0212026].

\bibitem{Meissner2004}
U. -G. Mei{\ss}ner, U. Raha, and A. Rusetsky, Eur. Phys. J. C {\bf 35} (2004) 349
[arXiv:hep-ph/0402261].

\bibitem{Ivanov2004}
  A.~N.~Ivanov, M.~Cargnelli, M.~Faber, J.~Marton, N.I. Troitskaya and J. Zmeskal,
  Eur. Phys. J. A {\bf 21} (2004) 11
  [nucl-th/0310081].

\bibitem{Ivanov2005}
A. N. Ivanov, M. Cargnelli, M. Faber, H. Fuhrmann, V. A. Ivanova, J. Marton, N. I. Troitskaya and J. Zmeskal,
Eur. Phys. J. A {\bf 25}, (2005) 329
[nucl-th/0505078].

\bibitem{Borasoy2005}
  B.~Borasoy, R.~Ni{\ss}ler and W.~Weise,
Phys.\ Rev.\ Lett.\  {\bf 94} (2005) 213401
  [arXiv:hep-ph/0410305];
  Eur.\ Phys.\ J.\ A {\bf 25} (2005) 79
  [arXiv:hep-ph/0505239].


\bibitem{Oller2005}
  J.~A.~Oller, J.~Prades and M.~Verbeni,
  Phys.\ Rev.\ Lett.\  {\bf 95} (2005) 172502
  [arXiv:hep-ph/0508081].

\bibitem{Borasoy2006}
  B.~Borasoy, U.-G.~Mei{\ss}ner and R.~Ni{\ss}ler,
   Phys.\ Rev.\ C {\bf 74} (2006) 055201
  [arXiv:hep-ph/0606108].

\bibitem{Trueman}
T.L. Trueman,
Nucl. Phys. {\bf 26} (1961) 57.

\bibitem{Deser}
S. Deser, M.L. Goldberger, K. Baumann, W. Thirring,
Phys. Rev. {\bf 96} (1954) 774.

\bibitem{Auvil}
P.R. Auvil, Phys. Rev. D {\bf 4} (1971) 240.

\bibitem{Pearce}
B.C. Pearce, B.K. Jennings, Nucl. Phys. A {\bf 528} (1991) 655.

\bibitem{Gibbs}
W.R. Gibbs, Li Ai, W.B. Kaufmann, Phys. Rev. C {\bf 57} (1998) 784.

\bibitem{Gashi1}
A. Gashi, E. Matsinos, G. C. Oades, G. Rasche, W. S. Woolcock,
Nucl. Phys. A {\bf 686} (2001) 463
[arXiv:hep-ph/0009080].

\bibitem{Gashi2}
A. Gashi, G. Rasche, W.S. Woolcock,
Phys. Lett. B {\bf 513} (2001) 269
[arXiv:hep-ph/0104059].

\bibitem{Gashi3}
A. Gashi, G.C. Oades, G. Rasche, W.S. Woolcock,
Nucl. Phys. A {\bf 699} (2002) 732
[arXiv:hep-ph/0108116].

\bibitem{AY1}
Y. Akaishi, T. Yamazaki,
Phys. Rev. C {\bf 65} (2002) 044005.

\bibitem{AY2}
T. Yamazaki, Y. Akaishi,
Phys. Rev. C {\bf 76} (2007) 045201
[arXiv:nucl-th/0709.0630].

\bibitem{Hyodo}
T. Hyodo, W. Weise,
Phys. Rev. C {\bf 77} (2008) 035204
[arXiv:nucl-th/0712.1613].

\bibitem{Martin}
  A.~D.~Martin,
  Nucl.\ Phys.\ B {\bf 179} (1981) 33.

\bibitem{KEK}
M. Iwasaki {\it et al} [KEK Collaboration],
Phys. Rev. Lett. {\bf 78} (1997) 3067.

\bibitem{Amirk}
I. Amirkhanov, I. Puzynin, A. Tarasov, O. Voskresenskaya, O. Zeinalova,
Phys. Lett. B {\bf 452} (1999) 155.

\bibitem{Stur}
M. Rotenberg,
Adv. At. Mol. Phys. {\bf 6} (1970) 233.

\bibitem{Yanatom}
Y. Yan, R. Tegen, T. Gutsche, A. Faessler,
Phys. Rev. C {\bf 56} (1997) 1596.

\bibitem{Yanpionium}
P. Suebka, Y. Yan,
Phys. Rev. C {\bf 70} (2004) 034006.






\end{thebibliography}
\end{document}